\documentclass[submission,copyright,creativecommons]{eptcs}
\providecommand{\event}{FROM 2025 (Working Formal Methods Symposium)}

\usepackage{iftex}
\usepackage{xcolor} 
\usepackage{enumitem}
\usepackage{graphicx}
\usepackage{scalerel}

\ifpdf
  \usepackage{underscore}         
  \usepackage[T1]{fontenc}        
\else
  \usepackage{breakurl}           
\fi

\title{GView: A Survey of Binary Forensics via Visual, Semantic, and AI-Enhanced Analysis}
\author{Raul Zaharia~\orcidicon{0009-0005-6366-9152}\qquad\qquad Dragoș Gavriluț~\orcidicon{0009-0004-3339-9625} \qquad\qquad Gheorghiță Mutu~\orcidicon{0009-0007-6998-9469}
\institute{Al. I. Cuza University \& Bitdefender, Iași, Romania}
\email{rzaharia@bitdefender.com}
}

\usepackage{tikz}
\usetikzlibrary{svg.path}
\def\BibTeX{{\rm B\kern-.05em{\sc i\kern-.025em b}\kern-.08em
    T\kern-.1667em\lower.7ex\hbox{E}\kern-.125emX}}

\definecolor{orcidlogocol}{HTML}{A6CE39}
\tikzset{
  orcidlogo/.pic={
    \fill[orcidlogocol] svg{M256,128c0,70.7-57.3,128-128,128C57.3,256,0,198.7,0,128C0,57.3,57.3,0,128,0C198.7,0,256,57.3,256,128z};
    \fill[white] svg{M86.3,186.2H70.9V79.1h15.4v48.4V186.2z}
                 svg{M108.9,79.1h41.6c39.6,0,57,28.3,57,53.6c0,27.5-21.5,53.6-56.8,53.6h-41.8V79.1z M124.3,172.4h24.5c34.9,0,42.9-26.5,42.9-39.7c0-21.5-13.7-39.7-43.7-39.7h-23.7V172.4z}
                 svg{M88.7,56.8c0,5.5-4.5,10.1-10.1,10.1c-5.6,0-10.1-4.6-10.1-10.1c0-5.6,4.5-10.1,10.1-10.1C84.2,46.7,88.7,51.3,88.7,56.8z};
  }
}

\newcommand\orcidicon[1]{\href{https://orcid.org/#1}{\mbox{\scalerel*{
\begin{tikzpicture}[yscale=-1,transform shape]
\pic{orcidlogo};
\end{tikzpicture}
}{|}}}}
\newcommand{\orcidlinktext}[1]{\url{https://orcid.org/#1}}

\def\titlerunning{GView: A Survey of Binary Forensics}
\def\authorrunning{R. Zaharia et al.}
\begin{document}
\maketitle

\begin{abstract}
Cybersecurity threats continue to become more sophisticated and diverse in their artifacts, boosting both their volume and complexity. To overcome those challenges, we present GView, an open-source forensic analysis framework with visual and AI-enhanced reasoning. It started with focus on the practical cybersecurity industry. It  has evolved significantly, incorporating large language models (LLMs) to dynamically enhance reasoning and ease the forensic workflows. This paper surveys both the current state of GView with its published papers and work that is currently under review. It also includes its innovative use of logical inference through predicates and inference rules for both the analyzed documents and the user's actions for better suggestions. We highlight the extensible architecture, showcasing its potential as a bridge between the practical forensics world with the academic research.
\end{abstract}

\section{Introduction}

The number of malicious files is continuously increasing, reaching over 1.2 billion and almost 290,000 new malware and PUA (possibly unwanted software) per day~\cite{av-test-malware-stats}. This continuous malicious development increases both the number and also the complexity of the attacks~\cite{StayingAheadOfThreatActors}. Modern attack vectors rely on different artifacts written in multiple programming languages capable of targeting any OS. To fully comprehend each step of the attack, a forensics analyst must understand every piece of information and extract the most relevant data in a timely manner. This is a difficult task since it often requires knowing a large variety of tools, from understanding binary data to extracting and sending them for re-analysis. The task is even more costly when the analyst needs to export some intermediary artifacts to be analyzed in another tools, making their job even more difficult. This fragmented workflow slows down the investigation and increases the risk of human error and analytic blind spots.

To address these limitations, we developed GView (Generic View)~\cite{gview-softwarex}\cite{gview-scid}, an open-source framework for interactive forensic analysis. GView is not just a viewer, but a reasoning-driven investigative environment built to support complex analysis tasks. Its design was formed on four scientific challenges (SC), each rooted in distinct forms of reasoning:

\begin{itemize}[itemsep=1pt,topsep=2pt,partopsep=1pt]
\item [\textbf{SC1}] \textbf{Automated Extraction and Understanding}
GView automatically identifies and semantically extracts content from 40+ binary formats. This involves structural reasoning using grammar and context-sensitive heuristics to prioritize relevant artifacts that can be re-analyzed inside the same tool.

\item [\textbf{SC2}] \textbf{Correlation and Inference of Forensic Hints}
Through inductive reasoning, GView helps the analyst to gain an overview of partially observed data by carefully presenting the most relevant information. This enables users to build logical chains of inference and surface hidden connections between disparate components.


\item [\textbf{SC3}] \textbf{Multi-Granularity Visual Reasoning} GView presents various visualizations, from byte-level to high-level, enabling analysts to pinpoint and comprehend the steps present in the data. They can form a conclusion based on both the current artifact and the entire analysis of all the steps taken.

\item [\textbf{SC4}] \textbf{Expert-System Guidance and Decision Support}
GView aims to become an expert system that actively supports users throughout their analysis. Its goal is to observe how users do their work, recognize various patterns (in both their actions and in the analyzed content), and suggest logical next steps. This built-in reasoning helps guide less experienced analysts while also making the investigation more efficient for experts. 

\item [\textbf{SC5}] \textbf{Extensibility and Future-Readiness
}
GView is designed with a modular architecture that can be easily updated to follow the latest analytical trends to keep up with the evolution of malicious files. Unlike traditional tools that depend on hand-crafted rules and actions, GView is built to be flexible and future-ready, especially through its integration with large language models.

\end{itemize}

\section{Architecture and Design of GView}

\begin{figure}[t]
\vspace{-2ex}
\centerline{\includegraphics[width=0.9\linewidth]{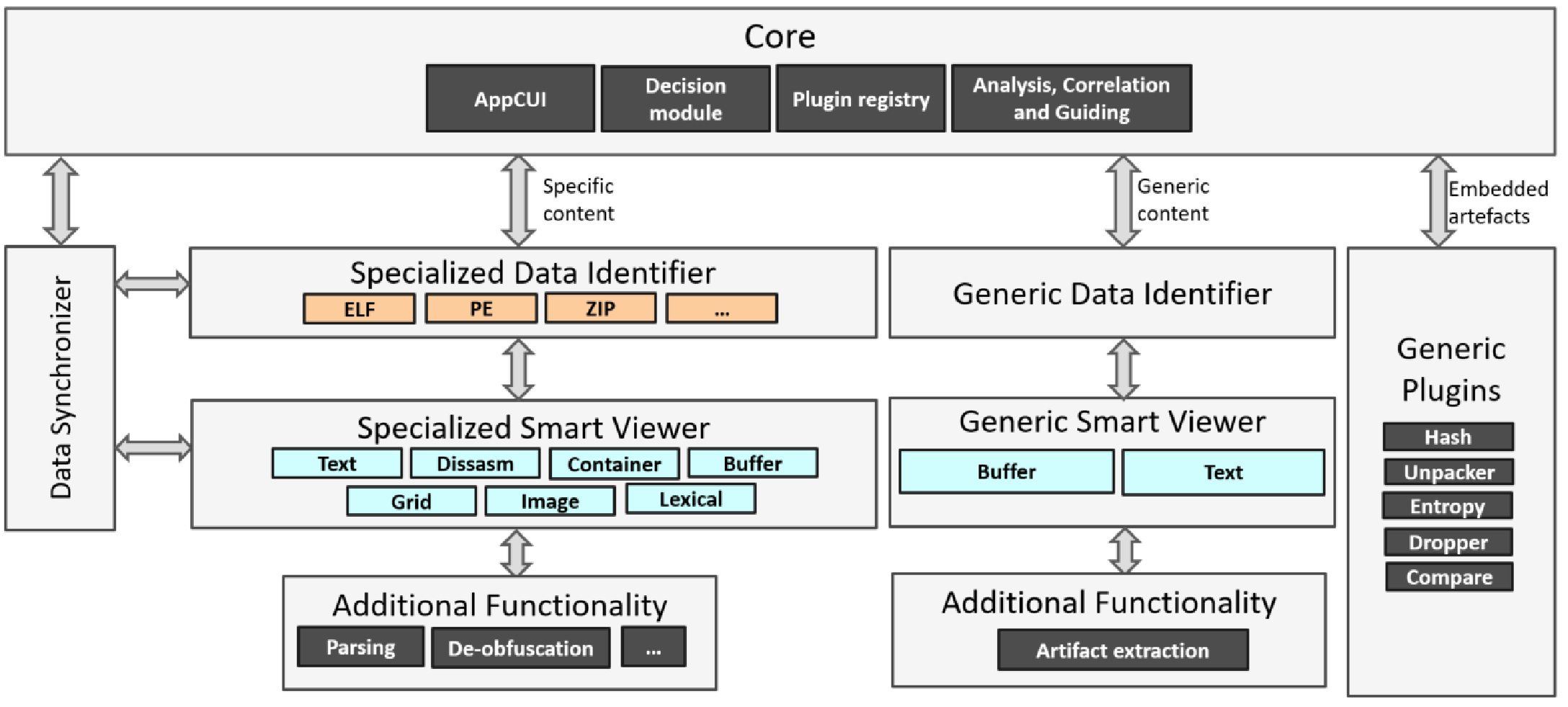}}
\vspace{-2.5ex}
\caption{GView architectural design}
\vspace{-3ex}
\label{fig:gview_architecture}
\end{figure}

The architecture of GView is designed to follow the Plug-and-Play pattern~\cite{Birsan05} enabling seamless integration and use of new components (see Figure~\ref{fig:gview_architecture}). Around this core, GView offers modular plugins into two primary types:
\begin{itemize}
    \item \textbf{Data Identifiers} are responsible for analyzing content and extracting the most meaningful information based on data type.
    \item \textbf{Smart Viewers} showcase the extracted information using the most relevant visualization type for that kind of data.
\end{itemize}

Plugins can be specialized or generic. Specialized plugins focus on content-specific tasks such as parsing using specific file formats, refactoring, and de-obfuscation. Generic plugins handle broader tasks, such as artifact extraction, structural analysis of unknown data types.

\subsection{Functional Overview}
GView uses a content-based analysis strategy. Upon receiving content, the core engine checks with each of the data identifier plugins if they can understand the content. If there are no specific plugins for that type of content, then a generic one is chosen. The retrieved information is then sent to or more relevant smart viewers, each capable of rendering different perspectives of the content. If some parts are of interest (intermediary artifacts), the user can select those for reanalysis.

The entire analysis process is both iterative (same steps are repeated) and dynamic. Artifacts extracted in one step may contain additional content that need further processing (e.g, unpacking or de-obfuscation) that GView can do. To be able to work efficiently with the data, GView introduces two new methodologies: a collaborative mechanism, where data identifiers exchange information with smart viewers, and a guiding mechanism to help the analyst.

\subsection{Guided and Collaborative Analysis}
GView's design extends beyond traditional plug-and-play limits by introducing an integrated \textbf{collaboration mechanism} that correlates the extracted data across analysis steps. This means that artifacts which usually require multiple analysis steps and tools can be easily done in GView, using the available plugins and viewers.

This is a reasoning-driven workflow where each analysis step builds upon the previous one, where GView not only extracts the most relevant information but also interprets it and offers suggestions (e.g., malicious indicators, irregularities in some sections). The recursive analysis of data, until every byte is understood, improves the analyst's level of confidence in their verdict. 

Complementing this is \textbf{guided security analysis} which highlights malicious patterns in the extracted information. Data identifiers flag suspicious constructs and delegate post-processing operations (e.g., de-obfuscation, code unpacking, etc.) to associated plugins. This guidance system helps the analyst to focus on the most important areas of the data.

\section{Use Case}
In this section, we examine how GView can assist a forensics analyst in two real-world inspired scenarios, where we assess the malicious behavior. In the first one, we try to understand how the user got infected, starting from the network traffic. On the second one, we start with the initial phishing email.

\subsection{Network-Based Infection Analysis}


We evaluate GView in a simulated yet realistic ransomware attack to demonstrate how it supports reasoning-driven forensic analysis. The attack begins when the user, Bob, attempts to download Firefox from a phishing site, leading to the execution of a password-protected, self-extracting ransomware.

From a reasoning perspective, this scenario requires multistep deduction and artifact correlation (usually done with multiple tools):
\begin{enumerate}
    \item \textit{Traffic analysis}: Upon opening the file that contains the network traffic, GView showcases the connections, and we find one that downloads a suspicious JavaScript file. We download that file for further analysis.
    \item \textit{Semantic code analysis}: The file is heavily obfuscated and challenging to read. After some de-obfuscation actions and some reasoning over syntax and control flow, we recover two hidden payload URLs. One URL is points to an image that have some text inside it, resembling a password (easily analyzed inside GView). The other one is leads to an executable.
    \item \textit{Executable inspection}: Upon further inspection, the executable is actually a self-extracting executable, meaning that it is some other data hidden inside it. That data is actually a password protected archive (the password from the image works here). We obtain another executable. 
    \item \textit{Disassembly and intent inference}: using the DissasmViewer (smart viewer helps the analyst to look at the disassembled code~\cite{dissasm-synasc}) we observe that this application tries to appear to be a Word document (trying to use its icon and description). We also identify the ransomware logic and ransom note encoding. We identified the ASCII Art obfuscation, decoded it, thus showing GView’s ability to reason over non-standard representations.
    \item \textit{Cross-artifact correlation}: Multiple malicious indicators are found: IP addresses, registry keys, wallet addresses. 
\end{enumerate}

GView integrates a reasoning workflow, helping the analyst to take decisions and follow a direction. Its strength lies in the logic that drives the forensic discovery. 

\subsection{Suspicious Email Dissection}


In this scenario, we showcase how GView deals with a case inspirited by real-world tactics~\cite{hackers-sponsorship-pdf}: how a phishing attack masking itself as a sponsorship email~\cite{email-gview-screencast}.

The attack starts when the user, Bob, receives an email containing a password-protected archive and is instructed to open it using the details inside the email. The archive contains a malicious executable disguised as a PDF file, called \textit{Contracts.pdf.exe}. There are used two main techniques: double-extension masquerading and a deception icon resembling a PDF document.

The analysis process consists of:
\begin{enumerate}
    \item \textit{Decomposition of Email}: Upon opening the file where the initial email was saved (as an \textit{.eml} document), GView parses and extracts the email components: body, attachments, other metadata. We can easily follow the content and all the available details, such as the password and the password-protected archive. 
    \item \textit{Interpretation of the File Metadata}: The content from within the archive is extracted and opened using GView. We easily observe the double extension, and its icon. All these indicators lead to suspicion, possible malicious behavior. 
    \item \textit{Artifact Correlation and Threat Deduction}: GView identifies all the strings, file paths and resources (registry keys, wallet addresses) and marks those that often correspond to attacks (such as registry keys used in persistence). 
    \item \textit{Behavioral Inference via Disassembly}: GView disassembles the binary. It recognizes all the strings and presents which OS APIs are used and their parameters. Using this, we observe some keystroke tracking, inferring malicious behavior from the control flow semantics.
\end{enumerate}

GView makes the analysis easier by providing additional context, not just raw maliciousness indicators. Instead of using multiple tools (e.g., email parsers, archive extractors, disassemblers, etc.) GView offers all their functionalities in a single place, facilitating all the required functionality at hand. Furthermore, the analyst can easily review all the taken steps and all the intermediary artifacts.



\section{Enhancing GView with LLMs}

The continuous advancement of threats requires the tools used to also be kept up to date, and, sometimes, that is difficult to achieve. Furthermore, traditional static analysis tools might be lacking the exact feature required in a specific scenario. To address this, we explored how large language models (LLMs) could extend its capabilities and reasoning potential~\cite{llms-gview}.

We leveraged GView's plug-and-play architecture to add a new module (LLM) that communicates through interfaces, ensuring extensibility. Existing components (data identifiers and smart viewers) can easily communicate the new module.

The core now offers two reasoning-support functions:
\begin{itemize}
    \item \textit{Contextual Prompt Augmentation}: To increase the efficiency of the LLM result, GView provides an enriched context, taking advantage of its already existing components. This increases the LLM's ability to generate more accurate and meaningful inferences.
    \item \textit{Prompt Optimization}: GView tries to optimize the context to consider the most relevant available data based on semantics, to try to minimize the final cost while also receiving accurate results. For instance, when dealing with Portable Executable (PE — Windows binaries), GView has a lot of information about them, but not all that information helps the LLM, some even might hinder it by misleading to the wrong direction.
\end{itemize}

From the end user's perspective, two new easy-to-use functionalities are added:
\begin{itemize}
    \item \textit{Interactive Chat}: The analyst can query the LLM about the current analyzed document, and GView will add the optimized context, so the LLM can fulfill the request. This greatly reduces the analysis time since the user can ask the LLM directly if something is unclear or needs explanation instead of manually creating such a prompt. 
    \item \textit{Automated Analytical Support}: Some analysis tasks are essential to every investigation. We automated those by adding direct commands (hotkeys and menu actions) for doing them instead of manually doing that: asking the LLM to rename a function and applying that change, asking for code explanation and inserting that as comments and more.
\end{itemize}

In our experiments, we used generic, out-of-the-box LLMs. While they could provide reasoning challenges (related to domain specificity), we try to address that using the rich and optimized context given. We systematically experimented and iterated through multiple variants of prompts until we reached the current ones. Formal methods, such as deductive reasoning and symbolic computation, guided our prompt engineering process since they offer context-enriching capabilities.

\section{Vision: Interactive, Inference-Driven Forensic Assistance}
\label{sec:vision}

To further improve the guiding capabilities of GView, we propose a new component, an inference-driven forensic assistance engine~\cite{gview-logical-engine}. It utilizes logical programming: predicates and formal inference rules. We categorize two types of predicates: automatically obtained by GView upon analyzing an artifact (analysis predicates, e.g., IsWord(File1), HasMacros(File1)) and analyst's actions (behavioral predicates, e.g., Opened(File1), ViewedMacros(File1)). Those will be used to suggest: possible actions to do as the current user for the file, and possible next steps for the analysis.

To obtain such actions, we will use a logical engine such as Z3, Datalog or even the integrated LLMs. The rules will be evaluated automatically, and the user will be presented with a textual description. RZData~\cite{csedu25} is a tool that translates programming code to its visual representation as data structures. We want to do something similar, but with predicates, where all inferred rules are translated into textual representations. This component would significantly extend GView's adaptability, supporting analysis with robust and logical forensic reasoning.

\section{Conclusion and Future Work}

GView was created initially for the practical industry need of analyzing evolving cybersecurity threats, having limited academic connections. Its flexible and modular architecture allowed it to become a framework easily extensible where both industry and academic practitioners can test, evaluate and develop their solutions. GView now bridges those two worlds: industry need of practical solutions with the academic world. Numerous research papers begin leveraging its extensibility, such as: Entropy-Driven Visualization~\cite{EntropyDrivenVisualization}, Analyzing Microsoft Office-Based Email Attacks~\cite{MicrosoftOffice-BasedEmailAttacks}, and several ongoing projects.

Future work will explore integrations with specialized and fine-tuned LLMs, while also providing the optimized reasoning framework described in Section~\ref{sec:vision}. The advanced inference engine based on logical programming will greatly improve the experience of both expert and novice security analysts.

\section{Bibliography}

\bibliographystyle{eptcs}

\begin{thebibliography}{10}
\providecommand{\bibitemdeclare}[2]{}
\providecommand{\surnamestart}{}
\providecommand{\surnameend}{}
\providecommand{\urlprefix}{Available at }
\providecommand{\url}[1]{\texttt{#1}}
\providecommand{\href}[2]{\texttt{#2}}
\providecommand{\urlalt}[2]{\href{#1}{#2}}
\providecommand{\doi}[1]{doi:\urlalt{https://doi.org/#1}{#1}}
\providecommand{\eprint}[1]{arXiv:\urlalt{https://arxiv.org/abs/#1}{#1}}
\providecommand{\bibinfo}[2]{#2}

\bibitemdeclare{inproceedings}{EntropyDrivenVisualization}
\bibitem{EntropyDrivenVisualization}
\bibinfo{author}{Andrada-Livia \surnamestart Antoneac\surnameend},
  \bibinfo{author}{Gheorghita \surnamestart Mutu\surnameend} \&
  \bibinfo{author}{Dragos-Teodor \surnamestart Gavrilut\surnameend}
  (\bibinfo{year}{2024}): \emph{\bibinfo{title}{{ Entropy-Driven Visualization
  in GView: Unveiling the Unknown in Binary File Formats }}}.
\newblock In: {\slshape \bibinfo{booktitle}{2024 26th International Symposium
  on Symbolic and Numeric Algorithms for Scientific Computing (SYNASC)}},
  \bibinfo{publisher}{IEEE Computer Society}, \bibinfo{address}{Los Alamitos,
  CA, USA}, pp. \bibinfo{pages}{74--81}, \doi{10.1109/SYNASC65383.2024.00025}.

\bibitemdeclare{misc}{av-test-malware-stats}
\bibitem{av-test-malware-stats}
\bibinfo{author}{\surnamestart {AV-Test}\surnameend}:
  \emph{\bibinfo{title}{{AV-ATLAS}}}.
\newblock \bibinfo{howpublished}{\url{https://portal.av-atlas.org/malware}}.
\newblock \bibinfo{note}{Accessed: 2025-05-25}.

\bibitemdeclare{article}{Birsan05}
\bibitem{Birsan05}
\bibinfo{author}{Dorian \surnamestart Birsan\surnameend}
  (\bibinfo{year}{2005}): \emph{\bibinfo{title}{On plug-ins and extensible
  architectures}}.
\newblock {\slshape \bibinfo{journal}{{ACM} Queue}}
  \bibinfo{volume}{3}(\bibinfo{number}{2}), pp. \bibinfo{pages}{40--46},
  \doi{10.1145/1053331.1053345}.

\bibitemdeclare{misc}{StayingAheadOfThreatActors}
\bibitem{StayingAheadOfThreatActors}
\bibinfo{author}{Microsoft~Threat \surnamestart Intelligence\surnameend}:
  \emph{\bibinfo{title}{{Staying ahead of threat actors in the age of AI}}}.
\newblock
  \bibinfo{howpublished}{\url{https://www.microsoft.com/en-us/security/blog/2024/02/14/staying-ahead-of-threat-actors-in-the-age-of-ai/}}.
\newblock \bibinfo{note}{Accessed: 2025-05-28}.

\bibitemdeclare{misc}{hackers-sponsorship-pdf}
\bibitem{hackers-sponsorship-pdf}
\bibinfo{author}{Jay \surnamestart Peters\surnameend} (\bibinfo{year}{2024}):
  \emph{\bibinfo{title}{How hackers took over Linus Tech Tips}}.
\newblock
  \bibinfo{howpublished}{\url{https://www.theverge.com/2023/3/24/23654996/linus-tech-tips-channel-hack-session-token-elon-musk-crypto-scam}}.
\newblock \bibinfo{note}{Accessed: 2025-05-15}.

\bibitemdeclare{inproceedings}{MicrosoftOffice-BasedEmailAttacks}
\bibitem{MicrosoftOffice-BasedEmailAttacks}
\bibinfo{author}{Cosmin \surnamestart Turtureanu\surnameend},
  \bibinfo{author}{Gheorghita \surnamestart Mutu\surnameend} \&
  \bibinfo{author}{Dragos-Teodor \surnamestart Gavrilut\surnameend}
  (\bibinfo{year}{2024}): \emph{\bibinfo{title}{{ A Software Engineering
  Approach into Analyzing Microsoft Office-Based Email Attacks }}}.
\newblock In: {\slshape \bibinfo{booktitle}{2024 26th International Symposium
  on Symbolic and Numeric Algorithms for Scientific Computing (SYNASC)}},
  \bibinfo{publisher}{IEEE Computer Society}, \bibinfo{address}{Los Alamitos,
  CA, USA}, pp. \bibinfo{pages}{113--117},
  \doi{10.1109/SYNASC65383.2024.00031}.

\bibitemdeclare{article}{dissasm-synasc}
\bibitem{dissasm-synasc}
\bibinfo{author}{Raul~\surnamestart Zaharia\surnameend} (\bibinfo{year}{2024}):
  \emph{\bibinfo{title}{Concepts Involved in Creating an Interactive Viewer for
  Disassembly}}.
\newblock {\slshape \bibinfo{journal}{Proceedings - 2024 26th International
  Symposium on Symbolic and Numeric Algorithms for Scientific Computing, SYNASC
  2024}}, pp. \bibinfo{pages}{107--110}, \doi{10.1109/SYNASC65383.2024.00029}.

\bibitemdeclare{inproceedings}{gview-scid}
\bibitem{gview-scid}
\bibinfo{author}{Raul \surnamestart Zaharia\surnameend},
  \bibinfo{author}{Dragos \surnamestart Gavrilut\surnameend},
  \bibinfo{author}{Gheorghita \surnamestart Mutu\surnameend} \&
  \bibinfo{author}{Dorel \surnamestart Lucanu\surnameend}
  (\bibinfo{year}{2024}): \emph{\bibinfo{title}{Interactive Assistance in
  Malware Dissemination Detection and Analysis}}.
\newblock In: {\slshape \bibinfo{booktitle}{Proceedings of the 1st Workshop on
  Security-Centric Strategies for Combating Information Disorder}},
  \bibinfo{series}{SCID '24}, \bibinfo{publisher}{Association for Computing
  Machinery}, \bibinfo{address}{New York, NY, USA},
  \doi{10.1145/3660512.3665526}.

\bibitemdeclare{article}{gview-softwarex}
\bibitem{gview-softwarex}
\bibinfo{author}{Raul \surnamestart Zaharia\surnameend},
  \bibinfo{author}{Dragoş \surnamestart Gavriluţ\surnameend},
  \bibinfo{author}{Gheorghiţă \surnamestart Mutu\surnameend} \&
  \bibinfo{author}{Dorel \surnamestart Lucanu\surnameend}
  (\bibinfo{year}{2024}): \emph{\bibinfo{title}{GView: A versatile assistant
  for security researchers}}.
\newblock {\slshape \bibinfo{journal}{SoftwareX}} \bibinfo{volume}{28}, p.
  \bibinfo{pages}{101940}, \doi{10.1016/j.softx.2024.101940}.

\bibitemdeclare{conference}{csedu25}
\bibitem{csedu25}
\bibinfo{author}{Raul \surnamestart Zaharia\surnameend} \&
  \bibinfo{author}{Dragoș \surnamestart Gavriluț\surnameend}
  (\bibinfo{year}{2025}): \emph{\bibinfo{title}{From Syntax to Sketch:
  Visualizing Code for Enhanced Comprehension with Focus on Cybersecurity}}.
\newblock In: {\slshape \bibinfo{booktitle}{Proceedings of the 17th
  International Conference on Computer Supported Education - Volume 2: CSEDU}},
  \bibinfo{organization}{INSTICC}, \bibinfo{publisher}{SciTePress}, pp.
  \bibinfo{pages}{752--759}, \doi{10.5220/0013295000003932}.

\bibitemdeclare{unpublished}{llms-gview}
\bibitem{llms-gview}
\bibinfo{author}{Raul \surnamestart Zaharia\surnameend},
  \bibinfo{author}{Dragoș \surnamestart Gavriluț\surnameend} \&
  \bibinfo{author}{Dorel \surnamestart Lucanu\surnameend}
  (\bibinfo{year}{2025}): \emph{\bibinfo{title}{Bridging Static Analysis and
  AI: An LLM-Based Approach to MITRE ATT\&CK Technique Identification}}.
\newblock \bibinfo{note}{Not yet published}.

\bibitemdeclare{unpublished}{gview-logical-engine}
\bibitem{gview-logical-engine}
\bibinfo{author}{Raul \surnamestart Zaharia\surnameend},
  \bibinfo{author}{Dragoș \surnamestart Gavriluț\surnameend} \&
  \bibinfo{author}{Dorel \surnamestart Lucanu\surnameend}
  (\bibinfo{year}{2025}): \emph{\bibinfo{title}{Predicate-Based Forensic
  Reasoning: Augmenting Analyst Workflow with Interactive Logical Inference}}.
\newblock \bibinfo{note}{Not yet published}.

\bibitemdeclare{misc}{email-gview-screencast}
\bibitem{email-gview-screencast}
\bibinfo{author}{Raul \surnamestart Zaharia\surnameend},
  \bibinfo{author}{Dragoș \surnamestart Gavriluț\surnameend},
  \bibinfo{author}{Gheorghiță \surnamestart Mutu\surnameend} \&
  \bibinfo{author}{Dorel \surnamestart Lucanu\surnameend}:
  \emph{\bibinfo{title}{GView: email sponsorship scenario}}.
\newblock \bibinfo{howpublished}{\url{https://youtu.be/LpsvcgCkII8}}.
\newblock \bibinfo{note}{Accessed: 2025-05-16}.

\end{thebibliography}

\end{document}